\documentclass{iau}
\usepackage{graphicx}

\graphicspath{{figs/}}

\title[Session 3 - Origin and impact of magnetic fields in higher-mass stars with radiative outer layers ] %% give here short title %%
{Magnetic fields of Ap stars from  full Stokes vector spectropolarimetric observations}

\author[N. Rusomarov, O. Kochukhov \& N. Piskunov]   %% give here short author list %%
{N. Rusomarov,
 O. Kochukhov
 \and
 N. Piskunov}

\affiliation{Department of Physics and Astronomy, Uppsala University, \\ Box
516, SE-75120 Uppsala, Sweden}

\pubyear{2013}
\volume{302}  %% insert here IAU Symposium No.
\jname{Magnetic Fields Throughout Stellar Evolution}
\editors{M. Jardine, P. Petit \& H. Spruit, eds.}
\begin{document}

\maketitle

\begin{figure}[h!]
  \centering
  \includegraphics[width=0.85\textwidth]{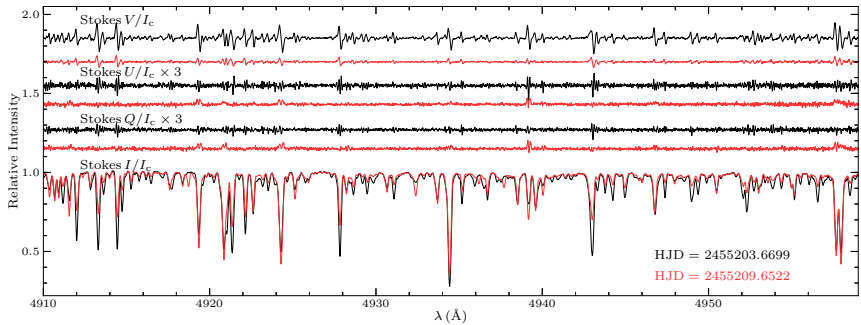}
  \caption{HARPSpol four Stokes parameter spectra of HD\,24712. Spectra plotted with black lines were obtained at magnetic maximum; spectra plotted with red lines were obtained around the magnetic minimum. Most lines exhibit strong intensity variations with phase, and show complex strong linear and circular polarization signatures.}
  \label{fig:order7_HD24712}
\end{figure}

%\section{Introduction}
Current knowledge about stellar magnetic fields relies almost entirely on circular polarization observations, with very few objects having been observed in all four Stokes parameters. We are investigating a sample of Ap stars in all four Stokes parameters using the HARPSpol instrument at the 3.6-m ESO telescope. In the context of this project we recently observed the magnetic Ap star HD\,24712 (DO Eri, HR\,1217). The resulting spectra have dense phase coverage, resolving power $> 10^5$, and S/N ratio of 300--600. These are the highest quality full Stokes observations obtained for any star other than the Sun. Furthermore, we have achieved good phase coverage for HD\,125248 and HD\,119419. Typical four Stokes parameters HARPSpol spectra are shown in Fig.~\ref{fig:order7_HD24712}. An analysis of the full Stokes vector spectropolarimetric data set of HD 24712 has been published in \cite{Rusomarov2013pA8}.

\begin{figure}
  \centering
  \includegraphics[width=0.45\textwidth, bb = 25 180 570 490, clip]{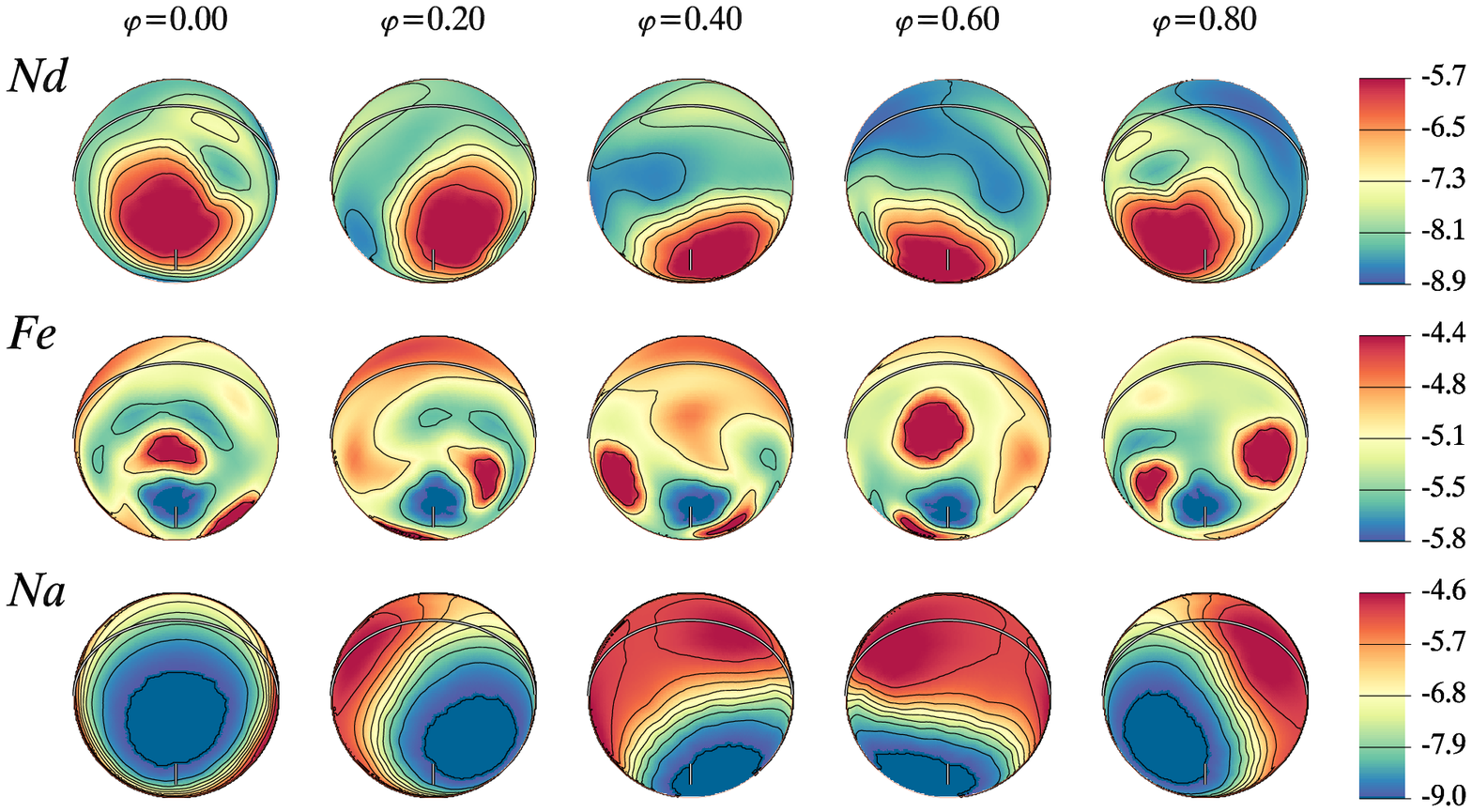} ~
  \includegraphics[width=0.45\textwidth, bb = 25 180 570 490, clip]{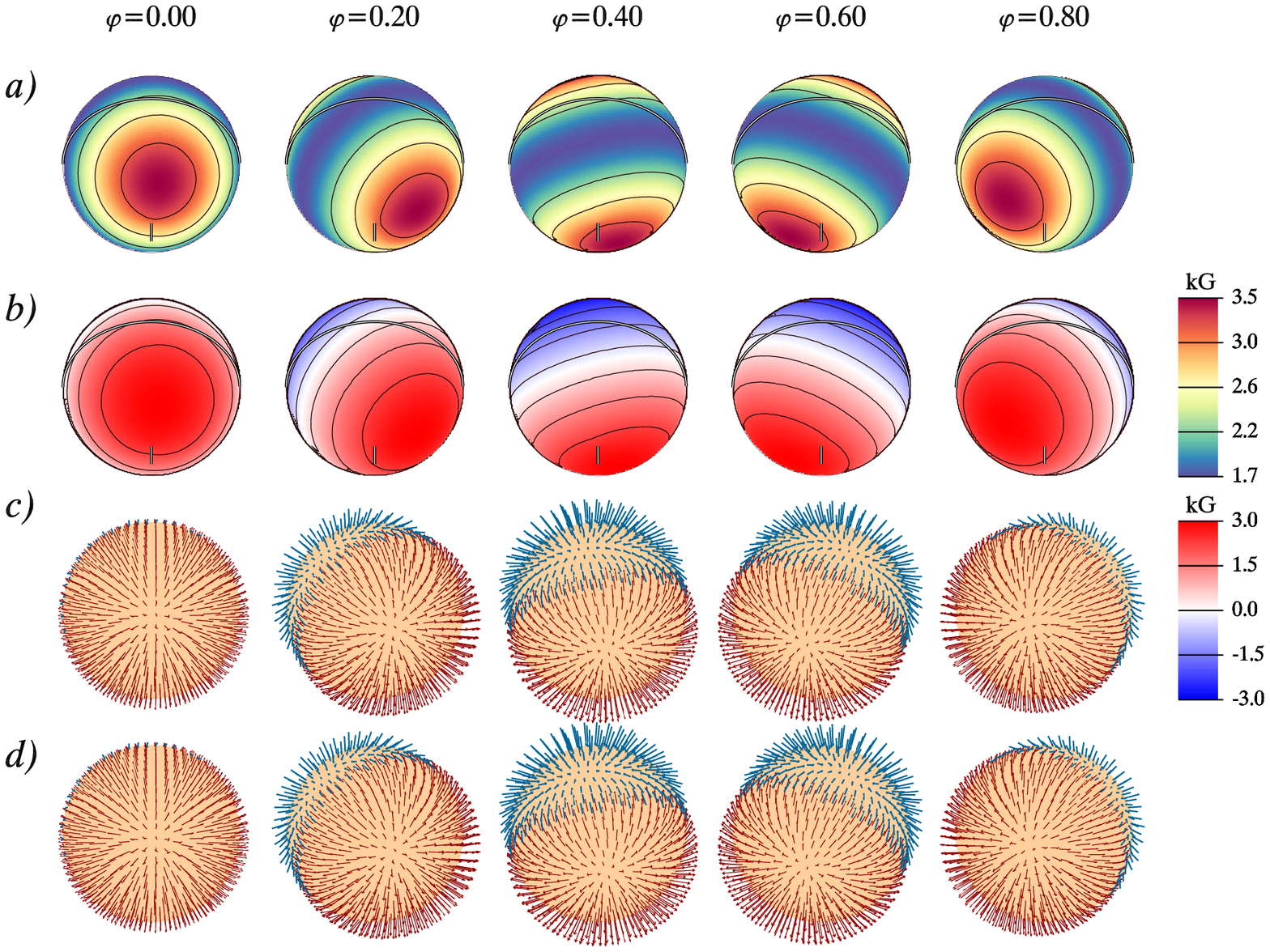}
  \caption{Abundance distribution maps of Nd, Fe and Na on the surface of HD24712, and maps showing a distribution of the magnetic field strength a), radial component b), and field orientation c).}
  \label{fig:MDI_HD24712_ab_mf}
\end{figure}

\begin{figure}
  \centering
  \includegraphics[width=0.95\textwidth, height=0.71\textheight, bb = 53 76 464 710, clip]{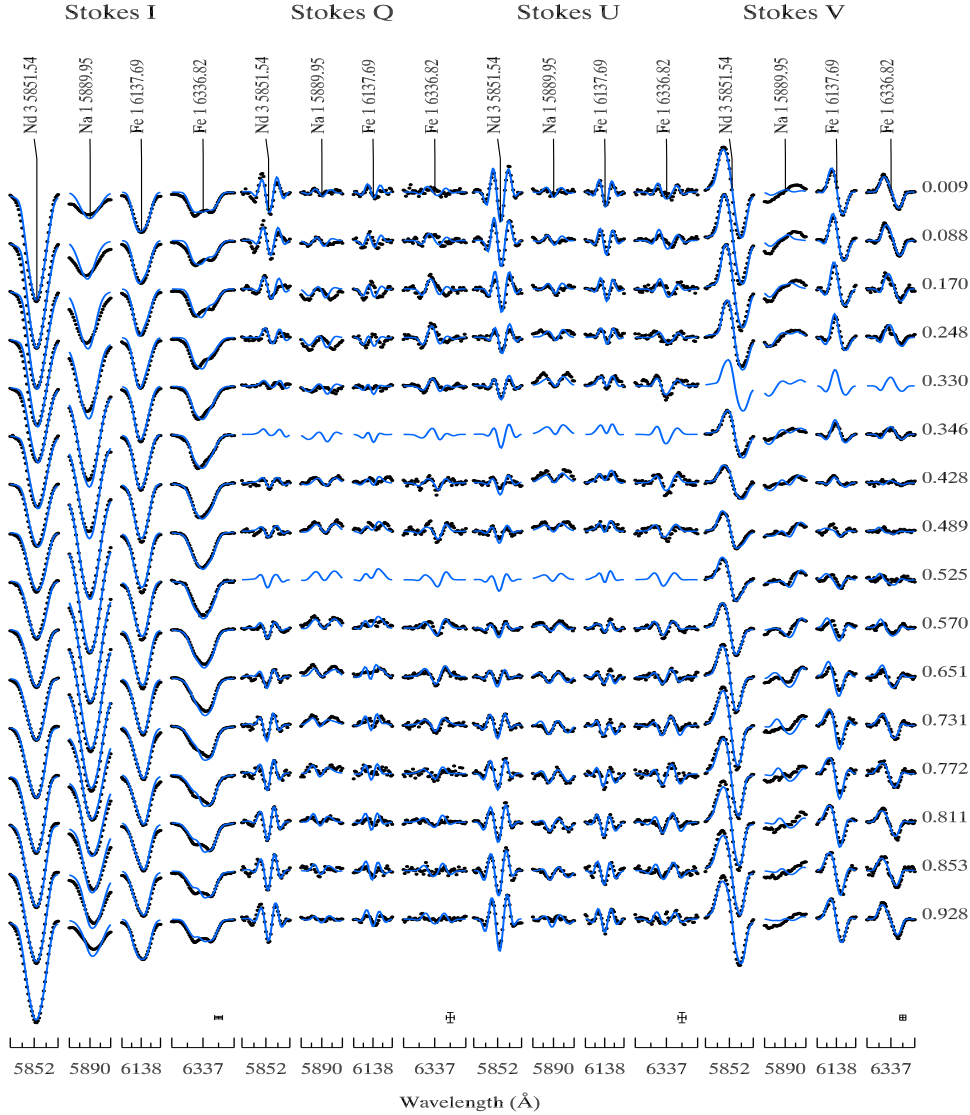}
  \caption{Comparison of the observed (dots) and calculated (lines) Stokes profiles for HD 24712. Phase values increase downwards.}
  \label{fig:MDI_HD24712_prof}
\end{figure}

Here we present preliminary results from the magnetic Doppler imaging analysis of HD\,24712. We derived chemical distribution abundance maps and magnetic field maps (Fig.~\ref{fig:MDI_HD24712_ab_mf}) using five FeI lines, three NdIII lines and one NaI line for the case of a dipolar field geometry. Our preliminary results show that dipole field geometry reproduces observed Stokes profiles for our selected lines very well (Fig.~\ref{fig:MDI_HD24712_prof}).

This analysis is the first step towards obtaining detailed 3-D maps of magnetic fields and abundance structures for HD\,24712 and other Ap stars that we currently observe with HARPSpol. We plan to study magnetic field and chemical spots in these stars, reconstruct 3D maps for the first time and analyze other stars.

\end{document}